\documentclass[twocolumn,printnumbers,amsmath,amssymb,prl]{revtex4}
\usepackage{graphicx}% Include figure files
\usepackage{color}

\begin{document}

\title{Jamming in confined geometry: Criticality of the jamming transition and implications of structural relaxation in confined supercooled liquids}

\author{Jun Liu$^1$}
\author{Hua Tong$^{2,\dagger}$}
\author{Yunhuan Nie$^1$}
\author{Ning Xu$^{1,*}$}

\affiliation{$^1$Hefei National Laboratory for Physical Sciences at the Microscale, CAS Key Laboratory of Microscale Magnetic Resonance and Department of Physics, University of Science and Technology of China, Hefei 230026, P. R. China\\
$^2$School of Physics and Astronomy, Shanghai Jiao Tong University, 800 Dong Chuan Road, Shanghai 200240, China}

\begin{abstract}
In marginally jammed solids confined by walls, we calculate the particle and ensemble averaged value of an order parameter, $\left<\Psi(r)\right>$, as a function of the distance to the wall, $r$. Being a microscopic indicator of structural disorder and particle mobility in solids, $\Psi$ is by definition the response of the mean square particle displacement to the increase of temperature in the harmonic approximation and can be directly calculated from the normal modes of vibration of the zero-temperature solids. We find that, in confined jammed solids, $\left<\Psi(r)\right>$ curves at different pressures can collapse onto the same master curve following a scaling function, indicating the criticality of the jamming transition. The scaling collapse suggests a diverging length scale and marginal instability at the jamming transition, which should be accessible to sophisticatedly designed experiments. Moreover, $\left<\Psi(r)\right>$ is found to be significantly suppressed when approaching the wall and anisotropic in directions perpendicular and parallel to the wall. This finding can be applied to understand the $r$-dependence and anisotropy of the structural relaxation in confined supercooled liquids, providing another example of understanding or predicting behaviors of supercooled liquids from the perspective of the zero-temperature amorphous solids.
\end{abstract}

\maketitle

\section{Introduction}
When being cooled down or compressed quickly, particles such as atoms, molecules, colloids, and grains can form amorphous solids via the noncrystalline liquid-solid transitions, e.g., the glass transition and the jamming transition \cite{debenedetti,liu_nagel}. The glass transition is signaled by the kinetic arrest of particles and the drastic growth of the viscosity with a small decrease of temperature or increase of pressure, whose nature remains one of the challenging puzzles in condensed matter physics \cite{berthier_rmp}. In contrast to the glass transition, which is normally for thermalized systems, the originally defined jamming transition is the formation of rigid packings of athermal particles when the system is compressed over a critical packing fraction $\phi_J$ \cite{ohern}. Typical features of the jamming transition include the isostaticity, i.e., the average coordination number per particle $z=z_c=2d$ at $\phi=\phi_J$ with $d$ being the dimension of space, and multiple critical scaling relations with respect to $\Delta \phi=\phi-\phi_J$ \cite{ohern,liu_review,hecke,xu_review1,xu_review2}. The study of marginally jammed solids, which are simplified zero-temperature model glasses at $\phi>\phi_J$, plays the crucial role in understanding the jamming transition and the peculiar features of supercooled liquids, even though the jamming transition is by definition distinct from the glass transition \cite{ikeda,parisi,charbonneau}.

Among various characteristics of marginally jammed solids, normal modes of vibration have exhibited their remarkable power to characterize the jamming transition and the glass transition. It has been shown that some characteristic frequencies of marginally jammed solids, e.g., the boson peak and Ioffe-Regel frequencies, decay to zero on approaching the unjamming transition, suggesting the existence of diverging length scales \cite{silbert,wyart1,wyart2,wyart3,xu1,wangxp,nie}. The short-time spatially heterogeneous dynamics in supercooled liquids have been shown to map well with the lowest-frequency modes of the corresponding inherent structures \cite{widmer,chenke}, i.e., zero-temperature states at local energy minima. Moreover, the boson peak and quasilocalization of low frequency modes of jammed solids can be employed to understand the density and interaction dependence of the glass transition and nonperturbative role of attraction in dynamics of supercooled liquids \cite{wanglj,berthier_attr,tonglj}.

The importance of normal modes of vibration can be further highlighted by a microscopic order parameter that we recently constructed, defined as the susceptibility of mean square particle displacement to infinitesimal thermal excitation in the zero-temperature limit \cite{tonghua_prl}:
\begin{equation}
\Psi_j=\frac{\partial \left<|\vec{u}_j^2|\right>}{\partial T},\label{eq:LDW}
\end{equation}
where $\vec{u}_j$ is the displacement of particle $j$ at a temperature $T$ with respect to its location at $T=0$, and $\left< .\right>$ denotes the time average.
In the harmonic approximation, $\Psi$ can be equivalently defined based on the assumption of energy equipartition under thermal excitation \cite{tonghua_pre}:
\begin{equation}
\Psi_j = \sum_{i=1}^{dN-d}\frac{1}{\omega_i^2}\left| \vec{e}_{i,j}\right|^2, \label{eq:psi}
\end{equation}
where $j$ is the particle index, $\omega_i$ is the frequency of the $i$-th mode, $\vec{e}_{i,j}$ is the polarization vector of particle $j$ in the $i$-th mode, $N$ is the total number of particles (rattlers, i.e., non-interacting particles, are removed from the analysis), and the sum is over all nontrivial normal modes. Therefore, $\Psi_j$ characterizes the mobility of particle $j$ under excitation and acts as a structural order parameter \cite{tonghua_pre,chenke_s2,chenke_psi,tong_nrp}: Particles with larger $\Psi$ are in more disordered structures and are easier to displace when excited. It has also been shown that $\Psi$ exhibits the power-law distribution and spatial correlation in marginally jammed solids, indicating the marginal stability of marginally jammed solids and the criticality of the jamming transition \cite{tonghua_prl}.

%Recently, we construct a microscopic order parameter from normal modes of vibration, based on the assumption of energy equipartition under thermal excitation \cite{tonghua_pre}:
%\begin{equation}
%\Psi_j = \sum_{i=1}^{dN-d}\frac{1}{\omega_i^2}\left| \vec{e}_{i,j}\right|^2, \label{eq:psi}
%end{equation}
%here $j$ is the particle index, $\omega_i$ is the frequency of the $i$-th mode, $\vec{e}_{i,j}$ is the polarization vector of particle $j$ in the $i$-th mode, \red{$N$ is the total number of particles rattlers, i.e., non-interacting particles, are removed from the analysis),} and the sum is over all nontrivial normal modes. In the harmonic approximation, \red{$\Psi$ can be equivalently defined as  \cite{tonghua_prl}}
%\begin{equation}
%\Psi_j=\frac{\partial \left<|\vec{u}_j^2|\right>}{\partial T},\label{eq:LDW}
%\end{equation}
%where $\vec{u}_j$ is the displacement of particle $j$ at a temperature $T$ with respect to its location at $T=0$, and $\left< .\right>$ denotes the time average. Therefore, $\Psi_j$ characterizes the mobility of particle $j$ under excitation and acts as a structural order parameter \cite{tonghua_pre,chenke_s2,chenke_psi,tong_nrp}: Particles with larger $\Psi$ are in more disordered structures and are easier to displace when excited. It has also been shown that $\Psi$ exhibits the power-law distribution and spatial correlation in marginally jammed solids, indicating the marginal stability of marginally jammed solids and the criticality of the jamming transition \cite{tonghua_prl}.

In this work, we move a step forward to utilize $\Psi$ to characterize the jamming criticality and relaxation dynamics of supercooled liquids. Different from most of previous studies of jamming, here we focus on systems confined by fixed walls. Intuitively, the presence of walls leads to spatial variations with the distance to the wall. The boundary effects are stronger closer to the wall, causing different behaviors from the bulk. For instance, it has been shown that the structural relaxation of confined supercooled liquids is slower near the wall \cite{watanabe,scheidler,cao} and the relaxation is anisotropic in the directions parallel and perpendicular to the wall \cite{hocky}. The order parameter $\Psi$ has shown its power to predict the short-time particle mobility. Can it work to predict the long-time structural relaxation of confined supercooled liquids? If then, it will be another paradigm of understanding supercooled liquids and glass transition from the perspective of the $T=0$ amorphous solids and demonstrate further the significance of $\Psi$. Moreover, the influence of the wall decays away from the wall. This decay naturally contains the information of some length scales. It is thus interesting to know whether such a length scale can be seen in confined marginally jammed solids and how it evolves on approaching the unjamming transition.

By calculating $\Psi$ in confined marginally jammed solids, we answer the above questions by the following findings. First, away from the wall, $\Psi$ monotonically increases, consistent with the behavior of the structural relaxation time of confined supercooled liquids. The vibration of low-frequency modes is significantly suppressed near the wall. Second, the component of $\Psi$ in the direction parallel to the wall is larger than the perpendicular one, consistent with previous observations that the structural relaxation time is shorter in the parallel direction. Third, $\Psi$ as a function of the distance to the wall exhibits the pressure dependence. Interestingly, all low-pressure data can collapse well onto a master curve following a scaling function. This scaling collapse suggests the existence of a length scale in marginally jammed solids, which diverges at the unjamming transition and thus characterizes the criticality of the jamming transition.

\section{Methods}

Here we focus on two-dimensional systems with a side length $L$. The system contains $N/2$ large and $N/2$ small disks with the same mass $m$ and a diameter ratio of $1.4$ to avoid crystallization. Particles interact via the pairwise harmonic potential:
\begin{equation}
U(r_{ij})=\frac{\epsilon}{2}\left( 1-\frac{r_{ij}}{\sigma_{ij}}\right)^2\Theta\left(1-\frac{r_{ij}}{\sigma_{ij}}\right),
\end{equation}
where $r_{ij}$ and $\sigma_{ij}$ are the separation between particles $i$ and $j$ and the sum of their radii, $\epsilon$ is the characteristic energy scale, and $\Theta(x)$ is the Heaviside step function. We set the units of mass, length, and energy to be $m$, small particle diameter $\sigma$, and $\epsilon$. The time is in units of $m^{1/2}\sigma\epsilon^{-1/2}$.

We first obtain marginally jammed solids at a given pressure $p$ with the periodic boundary conditions in both directions by quickly minimizing the enthalpy $H=U+pL^2$ of high-temperature states via the fast inertial relaxation engine algorithm \cite{fire}, where $U$ is the total potential energy of the system. For marginally jammed states of harmonic particles, $p\sim \Delta \phi$ \cite{ohern}, so $p$ also quantifies the distance to the unjamming transition.  We then divide the jammed solid into slabs with a thickness of $\sigma$ along the $y$-direction. A particle belongs to a slab when its center sits in the slab. Particles in the two slabs from $y=-\sigma$ to $y=\sigma$ are frozen into a wall, which are not allowed to move in all following calculations. Due to the periodic boundary conditions, the system is effectively confined by the wall from two sides. Figure~1 shows how the confined system is constructed. Because the wall is directly formed by particles in jammed states with periodic boundary conditions, this construction will not cause the interfacial stress mismatch between the wall and the movable particles.

%%%%%%%%%%%%%%%%%%%%%%%%%%%%%%%%%%%%%%%%%%%%%%%%%%%%%%%%%%%%%%%%%%%%%%%%%%
\begin{figure}
\includegraphics[width=0.4\textwidth]{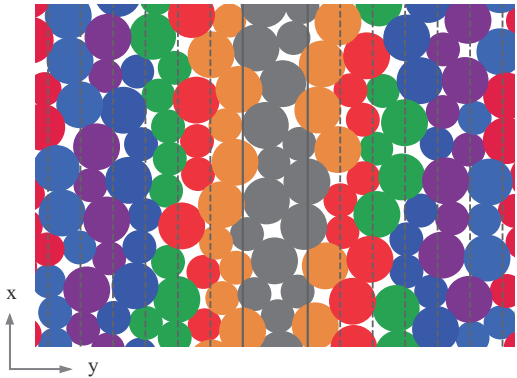}
  \caption{\label{fig:fig1} A central part of a marginally jammed state showing how the confined system is designed. The vertical solid lines at $y=\sigma$ and $y=-\sigma$ are boundaries of the wall, with gray particles in between being wall particles. The vertical dashed lines divide the system into layers with a thickness of $\sigma$. Particles in the same layer are displayed in the same color.}
\end{figure}
%%%%%%%%%%%%%%%%%%%%%%%%%%%%%%%%%%%%%%%%%%%%%%%%%%%%%%%%%%%%%%%%%%%%%%%%%%%

The normal modes of vibration are obtained by diagonalizing the Hessian matrix of the confined jammed solid. In the calculation of the Hessian matrix, the wall particles still interact with movable particles, but they do not contribute degrees of freedom, i.e., they are excluded in the calculation of $N$ in Eq.~(\ref{eq:psi}). In each slab, we calculate the average value of $\Psi$, denoted as $\left<\Psi\right>$. Then we obtain the profile $\left<\Psi(r)\right>$ averaged over thousands of independent jammed states, where $r$ is the distance of the slabs from the wall, which is $|y-\sigma|$ ($|y+\sigma|$) for $y>\sigma$ ($y<-\sigma$). Due to the periodic boundary conditions, $r$ cannot exceed $L/2$.

\section{Length scale and jamming criticality}

Figure~2(a) shows $\left<\Psi(r)\right>$ for different values of pressure $p$ and system size $N$. With the increase of $r$, $\left<\Psi\right>$ increases and seems to saturate on approaching $r=L/2$. This `saturation' is actually an artificial effect of finite system size and periodic boundary conditions. Note that the wall has two faces. In average, $\left<\Psi\right>(y)=\left<\Psi\right>(-y)$. The continuity of $\left<\Psi\right>$ at $y=\pm L/2$ leads to the `saturation'. At a given $p$, $\left<\Psi(r)\right>$ is almost independent of $N$ prior to the `saturation'. With the increase of $N$ and correspondingly $L(\sim N^{1/2})$, $\left<\Psi\right>$ goes to higher values. In the thermodynamics limit, the maximum value of $\left<\Psi\right>$ may be approximately that in the counterpart system without walls. From current data, it is hard to tell whether $\left<\Psi\right>$ approaches a finite value or diverges in the thermodynamic limit, which is beyond the scope of current study and will be discussed elsewhere.

Figure~2(a) also indicates that $\left<\Psi(r)\right>$ increases when pressure decreases. Surprisingly, Fig.~2(b) shows that all $\left<\Psi(r)\right>$ curves at different pressures can collapse nicely according to the scaling function
\begin{equation}
\left<\Psi(r)\right>=p^{-\alpha} f(rp^{\beta}), \label{eq:scaling}
\end{equation}
where $\alpha= 0.47\pm 0.05$ and $\beta= 0.25\pm 0.02$ are scaling exponents. Equation~(\ref{eq:scaling}) has a couple of implications. First, at the unjamming transition ($p=0$), $\left<\Psi(r)\right>$ diverges, demonstrating the marginal instability of the (un)jamming transition \cite{tonghua_prl}. Second, at a give $p$, there is a length scale $\lambda\sim p^{-\beta}\approx p^{-0.25}$, which diverges at the unjamming transition and suggesting the criticality of the jamming transition.

%%%%%%%%%%%%%%%%%%%%%%%%%%%%%%%%%%%%%%%%%%%%%%%%%%%%%%%%%%%%%%%%%%%%%%%%%%
\begin{figure}
\includegraphics[width=0.48\textwidth]{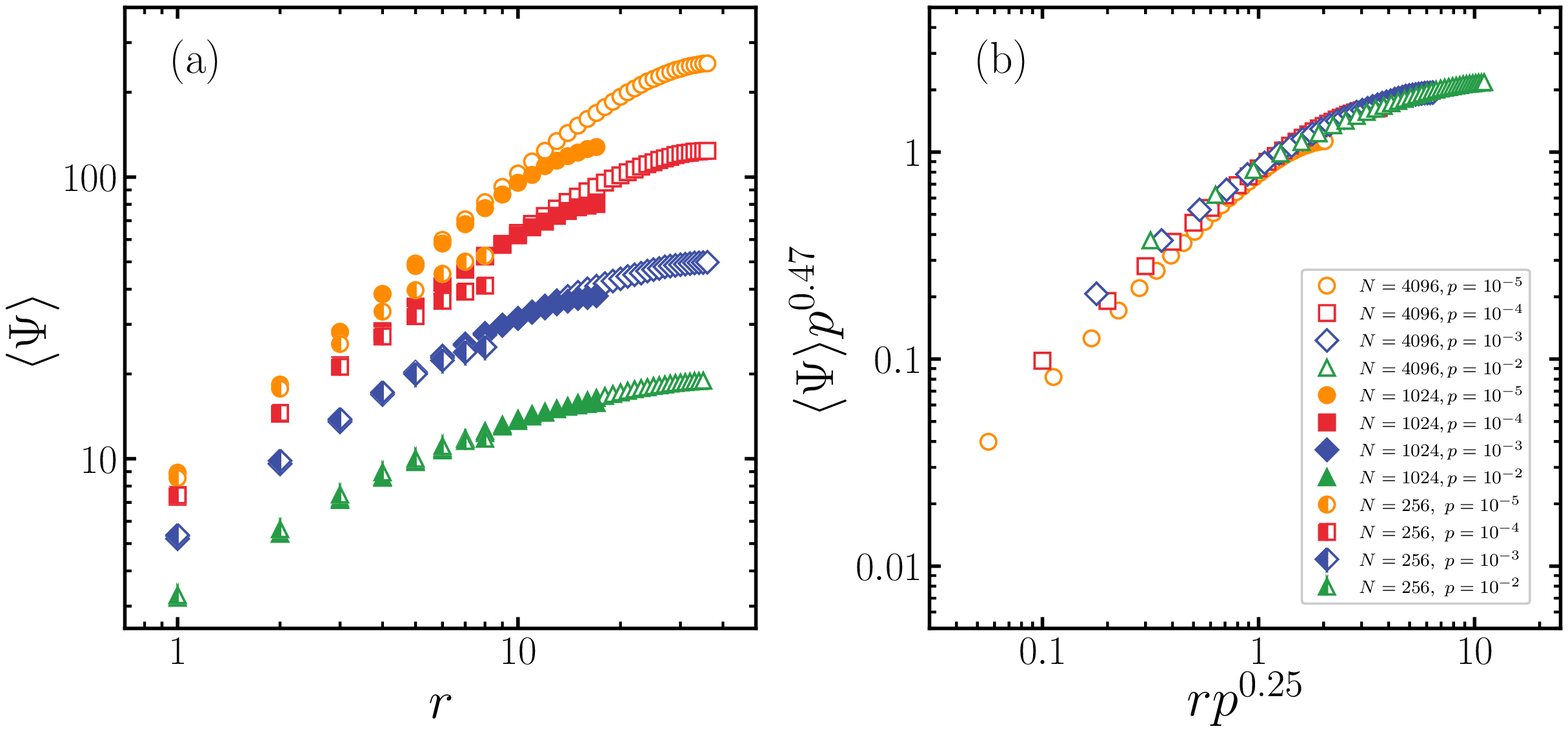}
  \caption{\label{fig:fig2} (a) System size and pressure dependence of the average order parameter $\left<\Psi\right>$ as a function of the distance to the wall, $r$. (b) Scaling collapse of $\left<\Psi(r)\right>$ curves for $N=4096$ systems at different pressures.}
\end{figure}
%%%%%%%%%%%%%%%%%%%%%%%%%%%%%%%%%%%%%%%%%%%%%%%%%%%%%%%%%%%%%%%%%%%%%%%%%%%

Our finding here suggests an experimentally accessible approach to directly probe the length scale in marginally jammed solids. As shown in Ref.~\cite{tonghua_pre}, $\Psi$ is by definition proportional to the local Debye-Waller factor in the harmonic approximation. When excited at sufficiently low temperatures, which do not cause particle rearrangements, particles in amorphous solids vibrate around their average locations. For particle $j$, $\left<|\vec{u}_j^2|\right>$ in Eq.~(\ref{eq:LDW}) actually measures its long-time mean square displacement, i.e., its Debye-Waller factor. Therefore, our approach here could be realized in sophisticated design of experimental systems such as colloidal suspensions.

There have been multiple reports of critical length scales in marginally jammed solids \cite{ohern,wangxp,xu1,silbert,olsson,ellenbroek,wanglj_jamming,wyart1,wyart2,wyart3,graves,liuhao,goodrich,liaoqy}. Is the length scale found here new or one of those previous reported? Seen from Eq.~(\ref{eq:psi}), $\Psi$ is directly calculated from the normal modes of vibration with a weight $1/\omega_i^2$, so lower frequency modes may contribute more to $\Psi$. As observed before, lowest-frequency modes of marginally jammed solids are mainly transverse and the characteristic frequencies of the transverse modes, e.g., the boson peak frequency and the transverse Ioffe-Regel frequency, decay to zero at the unjamming transition, implying the divergence of some length scales \cite{wangxp,wyart1,wyart2,wyart3,silbert}. It has been shown that such characteristic frequencies denoted as $\omega^*$ are scaled well with the pressure: $\omega^*\sim p^{1/2}$ \cite{wangxp,wyart1,wyart2,wyart3,silbert}. For harmonic systems, the transverse speed of sound is $c_T=\sqrt{G/\rho}$ with $G$ and $\rho$ being the shear modulus and mass density (which can be treated as a constant at low $p$ or $\Delta \phi$). It is well-known that $G\sim p^{1/2}$ for jammed solids of harmonic particles \cite{ohern}, so $c_T\sim p^{1/4}$. The dispersion relation thus leads to a length $\lambda_T\sim c_T/\omega^*\sim p^{-1/4}$. This is exactly the scaling of the length from Eq.~(\ref{eq:scaling}).

Equation~(\ref{eq:scaling}) also suggests that $\left<\Psi\right>\sim p^{-\alpha}$ when $rp^{\beta}$ remains constant. We will show next that this scaling is a direct manifestation of the significant role of $\omega^*$. Note that $\left<\Psi\right>$ is averaged over particles. If averaged over a sufficiently large number of particles or configurations, the average value of $|\vec{e}_{i,j}|^2$ in Eq.~(\ref{eq:psi}) will be approximately an $i$-independent constant, because the modes are normalized, so that
\begin{equation}
\left<\Psi\right>\sim \left<\sum_{i=1}^{dN-d}\frac{1}{\omega_i^2}\right>=\int_0^{\omega_m}\frac{1}{\omega^2}D(\omega){\rm d}\omega, \label{eq:psi_approx1}
\end{equation}
where $D(\omega)$ is the density of states, and $\omega_m$ is the maximum frequency of the modes. A very special feature of marginally jammed solids is that $D(\omega)$ is roughly split into two parts: When $\omega<\omega^*$, $D(\omega)\sim \omega^{\gamma}$ with $\gamma>0$, while $D(\omega)$ shows a plateau when $\omega^*<\omega<\omega_m$ \cite{silbert,wyart1,wyart2,wyart3,xu1}. The exponent $\gamma$ is not simply $d-1$ as predicted by the Debye law, which is an important issue under debate \cite{lerner1,lerner2,mizuno,charbonneau1,xu2}. Therefore, Eq.~(\ref{eq:psi_approx1}) turns to
\begin{equation}
\left<\Psi\right>\sim \int_0^{\omega^*}\frac{1}{\omega^2}D_0(p)\omega^{\gamma}{\rm d}\omega + \int_{\omega^*}^{\omega_m}\frac{1}{\omega^2}D_0(p)(\omega^*)^{\gamma}{\rm d}\omega, \label{eq:psi_approx2}
\end{equation}
where $D_0$ is a pressure-dependent prefactor of the density of states. For harmonic systems, it has been shown that the plateau value of $D(\omega)$ when $\omega>\omega^*$ is independent of pressure \cite{silbert,xu1}, so we can define a constant $D_c=D_0(p)(\omega^*)^{\gamma}$. Then Eq.~(\ref{eq:psi_approx2}) becomes
\begin{equation}
\left<\Psi\right>\sim \frac{D_c\gamma}{\gamma-1}(\omega^*)^{-1}-D_c\omega_m^{-1}. \label{eq:psi_approx3}
\end{equation}
For harmonic systems, $\omega_m$ is almost constant in pressure, so $\left<\Psi\right>\sim (\omega^{*})^{-1}\sim p^{-1/2}$ on approaching the unjamming transition, which is exactly the derivation from Eq.~(\ref{eq:scaling}). Note that Eq.~(\ref{eq:psi_approx3}) is derived from established knowledge of jammed solids without walls. Although the presence of walls changes the spatial organization of $\Psi$, the inherent properties of $\Psi$ should be maintained. Moreover, Eq.~(\ref{eq:psi_approx3}) requires that $\gamma>1$, which is consistent with previous studies. For two-dimensional harmonic systems, no scaling behaviors of the density of states with $\gamma<d-1=1$ have been seen at $\omega<\omega^*$. The Debye behavior with $\gamma=1$ for two-dimensional jammed states possibly exists at low frequencies, which may lead to some interesting consequences, such as the Mermin-Wagner fluctuation. If this is the case, the divergence of $\left<\Psi\right>$ suggested by Eq.~(\ref{eq:psi_approx3}) is a strong piece of evidence.

\section{Suppression of low-frequency vibration near the wall}

Figure~2(a) shows that $\left<\Psi\right>$ is small near the wall. From Eq.~(\ref{eq:psi}), this means that, at least for some modes, the polarization vectors of the particles near the wall are shorter than those away from the wall. It is straightforward to imagine that the wall constrains the motion of particles, especially in the direction perpendicular to it, and hence suppresses the mobilities of particles close to it. The question is in which modes the polarization vectors of particles are strongly suppressed near the wall. As discussed earlier, lower frequency modes should contribute more to $\Psi$, which naturally become the candidates.

To quantify the contribution of modes to the suppression, we define a parameter for mode $i$ with a frequency $\omega_i$:
\begin{equation}
h(\omega_i)=\sum_j \left|y_j\right|\left| \vec{e}_{i,j}\right|^2, \label{eq:h}
\end{equation}
where the sum is over all particles outside of the wall. If the polarization vectors of all particles have similar lengths, which is in fact a feature of extended modes in the intermediate frequency regime above $\omega^*$ \cite{xu1}, it is easy to see that $h\approx L/4$. For localized modes, e.g., high-frequency modes near $\omega_m$, $h$ will be either much larger or much smaller than $L/4$, depending on whether the localization of the vibration is far way from or near the wall. For low-frequency modes, which could be either phonon-like or quasilocalized (i.e., hybridization of phonon-like and localized) \cite{xu1,xu3}, $h$ should be close to $L/4$ as well. However, if the polarization vectors of particles close to the wall are suppressed, $h$ of these low-frequency modes will still be apparently larger than $L/4$.

%%%%%%%%%%%%%%%%%%%%%%%%%%%%%%%%%%%%%%%%%%%%%%%%%%%%%%%%%%%%%%%%%%%%%%%%%%
\begin{figure}
\includegraphics[width=0.48\textwidth]{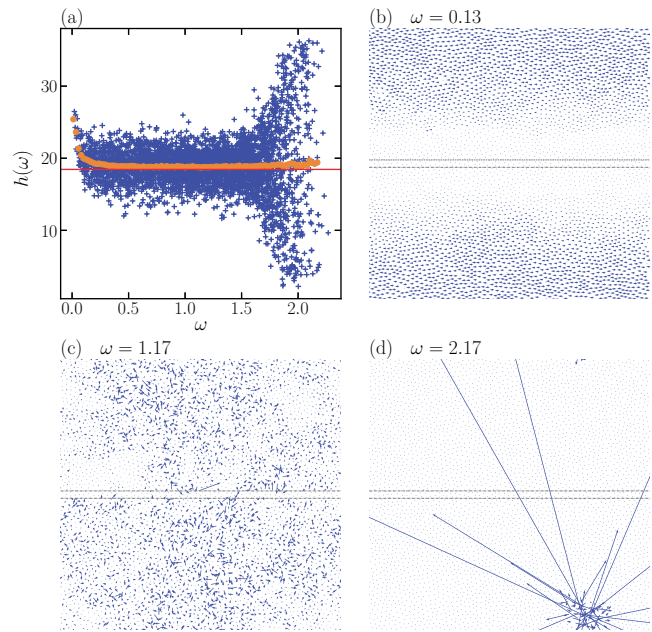}
  \caption{\label{fig:fig43} (a) Plots of $h(\omega)$ as defined in Eq.~(\ref{eq:h}) for $N=4096$ systems at $\phi=0.88$. The circles are averaged over tens of independent jammed states. The crosses show the scattered plot for a single state. The horizontal solid line shows $h=L/4$. (b)-(d) Polarization vector field of a mode with a frequency being shown on top. The horizontal dashed lines show boundaries of the wall. The direction and length of an arrow show the direction and magnitude of a polarization vector.}
\end{figure}
%%%%%%%%%%%%%%%%%%%%%%%%%%%%%%%%%%%%%%%%%%%%%%%%%%%%%%%%%%%%%%%%%%%%%%%%%%%

Figure~3(a) shows $h(\omega)$ averaged over tens of independent jammed states together with the scattered plot for one of the states. Apparently, for the lowest-frequency modes, $h$ is significantly larger than $L/4$ and shoots up with the decrease of $\omega$, which are more pronounced in the averaged $h(\omega)$. The behavior of $h(\omega)$ in the intermediate and high frequency regimes are as expected. Therefore, the presence of the wall indeed suppresses the low-frequency vibrations near the wall. This can be better visualized by the polarization vector fields shown in Figs.~3(b), (c), and (d). The particles near the wall are almost not involved in the low-frequency vibration (illustrated by the apparently short arrows near the wall in Fig.~3(b)), while the extended modes in the intermediate frequency regime (Fig.~3(c)) and high-frequency localized modes (manifested by the very long arrows in Fig.~3(d), indicating that only those localized particles with long arrows are effectively involved in the vibration) still look similar to those of systems without the wall, seemingly unaffected by the wall.

\section{Implications of structural relaxation in confined supercooled liquids}

Because $\Psi$ characterizes the mobility of particles \cite{tonghua_pre}, the fact that $\left<\Psi\right>$ increases away from the wall as shown in Fig.~2(a) indicates that particles near the wall are more difficult to move and cause structural relaxation. It would be expected that the structural relaxation time $\tau$ is longer on approaching the wall, which is exactly observed in confined supercooled liquids \cite{watanabe,scheidler,cao}. Therefore, in addition to the jamming criticality, the behavior of $\left<\Psi(r)\right>$ also sheds light on the structural relaxation in confined supercooled liquids. This provides one more piece of evidence indicating that behaviors of supercooled liquids can be understood or predicted from the perspective of $T=0$ amorphous solids, and hence underlines further the significance of the order parameter $\Psi$.

%%%%%%%%%%%%%%%%%%%%%%%%%%%%%%%%%%%%%%%%%%%%%%%%%%%%%%%%%%%%%%%%%%%%%%%%%%
\begin{figure}
\includegraphics[width=0.48\textwidth]{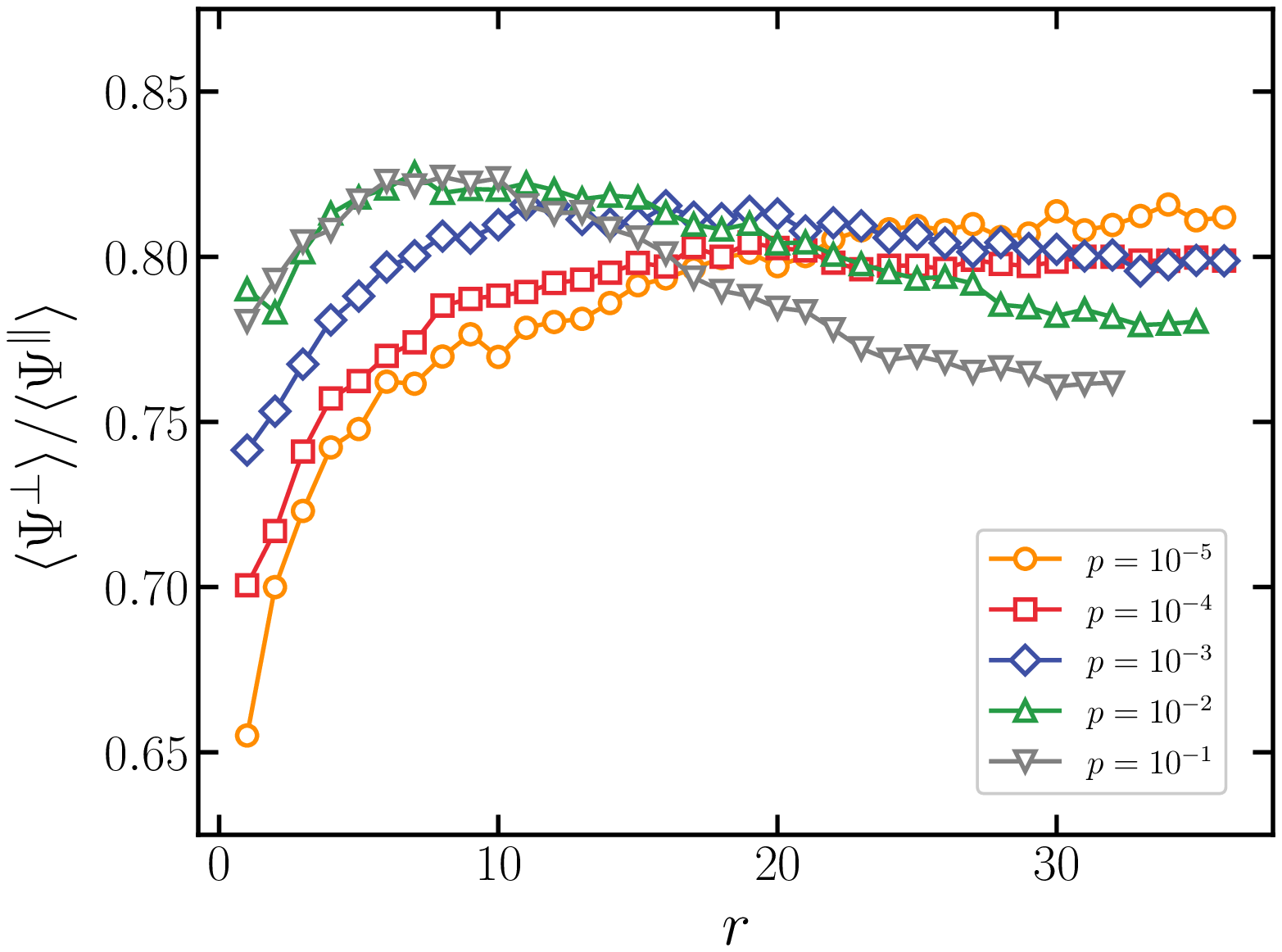}
  \caption{\label{fig:fig4} Anisotropy of $\left<\Psi(r)\right>$ quantified by $\left<\Psi^{\perp}(r)\right>/\left<\Psi^{\parallel}(r)\right>$ for $N=4096$ systems at different pressures.}
\end{figure}
%%%%%%%%%%%%%%%%%%%%%%%%%%%%%%%%%%%%%%%%%%%%%%%%%%%%%%%%%%%%%%%%%%%%%%%%%%%

Another special feature of confined supercooled liquids is that the structural relaxation is anisotropic. It has been found that the structural relaxation time $\tau^{\perp}$ perpendicular to the wall is larger than the parallel one $\tau^{\parallel}$, and the ratio of them shows peculiar $r$ dependence \cite{hocky}. These behaviors are understandable near the wall, because the wall impedes the motion of particles perpendicular to it. Away from the wall, the ratio $\tau^{\parallel}/\tau^{\perp}$ increases. For supercooled liquids at higher temperatures, $\tau^{\parallel}/\tau^{\perp}$ tends to approach a plateau close to $1$. With the decrease of temperature, the plateau value decreases. At even lower temperatures, the plateau disappears. Instead, away from the wall, $\tau^{\parallel}/\tau^{\perp}$ first increases and then decreases after reaching a maximum. Since we have seen the consistency between $\left<\Psi(r)\right>$ and $\tau(r)$, can we see any sign of the anisotropy from $\Psi$ as well?

We thus decompose $\Psi_j$ in Eq.~(\ref{eq:psi}) into two components:
\begin{equation}
\Psi_j^{\perp} = \sum_{i=1}^{dN-d}\frac{1}{\omega_i^2}\left| \vec{e}_{i,j}^{\perp}\right|^2, \label{eq:psi_perp}
\end{equation}
and
\begin{equation}
\Psi_j^{\parallel} = \sum_{i=1}^{dN-d}\frac{1}{\omega_i^2}\left| \vec{e}_{i,j}^{\parallel}\right|^2, \label{eq:psi_perp}
\end{equation}
where $\vec{e}_{i,j}^{\perp}$ and $\vec{e}_{i,j}^{\parallel}$ are the components of the polarization vector $\vec{e}_{i,j}$ perpendicular and parallel to the wall, respectively. Apparently, $\Psi_j=\Psi_j^{\perp}+\Psi_j^{\parallel}$. Because a larger $\Psi$ corresponds to a smaller $\tau$, is it possible that $\left<\Psi^{\perp}\right>/\left<\Psi^{\parallel}\right>$ behaves similarly to $\tau^{\parallel}/\tau^{\perp}$ as a function of $r$?

Figure~4 shows $\left<\Psi^{\perp}(r)\right>/\left<\Psi^{\parallel}(r)\right>$ for different pressures. At all pressures, $\left<\Psi^{\perp}(r)\right><\left<\Psi^{\parallel}(r)\right>$, so particle mobilities in the direction parallel to the wall are larger than those in the perpendicular direction. Following the correlation between $\Psi$ and $\tau$, it can be predicted that $\tau^{\parallel}/\tau^{\perp}<1$, which is exactly the case found before \cite{hocky}.

The shapes of $\left<\Psi^{\perp}(r)\right>/\left<\Psi^{\parallel}(r)\right>$ curves in Fig.~4 resemble those of $\tau^{\parallel}(r)/\tau^{\perp}(r)$, with the pressure dependence of $\left<\Psi^{\perp}(r)\right>/\left<\Psi^{\parallel}(r)\right>$ being like the temperature dependence of $\tau^{\parallel}(r)/\tau^{\perp}(r)$ mentioned above \cite{hocky}. At low pressures, $\left<\Psi^{\perp}(r)\right>/\left<\Psi^{\parallel}(r)\right>$ exhibits a plateau far away from the wall, similar to $\tau^{\parallel}(r)/\tau^{\perp}(r)$ at high temperatures. With the increase of pressure, a maximum emerges in $\left<\Psi^{\perp}(r)\right>/\left<\Psi^{\parallel}(r)\right>$, which is exactly what $\tau^{\parallel}(r)/\tau^{\perp}(r)$ looks like at low temperatures. Note that jammed solids are more stable with the increase of pressure, while supercooled liquids at lower temperatures also explore more stable metastable states. Therefore, $\Psi$ not only captures the anisotropic feature of the structural relaxation in confined supercooled liquids, but also helps to understand the temperature dependence of $\tau^{\parallel}(r)/\tau^{\perp}(r)$ from the perspective of stability. We have also verified that the shape of $\left<\Psi^{\perp}(r)\right>/\left<\Psi^{\parallel}(r)\right>$ does not show strong system size dependence, so it is expected that $\left<\Psi^{\perp}(r)\right>/\left<\Psi^{\parallel}(r)\right>$ will remain less than one even in the thermodynamic limit.

\section{Discussion and conclusions}

In summary, we characterize the behaviors of the order parameter $\left<\Psi\right>$ as a function of the distance to the wall, $r$, for confined marginally jammed solids on approaching the unjamming transition. The presence of the wall greatly suppresses the values of $\left<\Psi\right>$ close to it. Interestingly, $\left<\Psi(r)\right>$ curves at different pressures satisfy a universal scaling function, suggesting the criticality of the jamming transition. The scaling function implies a critical length scale, which diverges at the unjamming transition. The length scale agrees well with that from the transverse normal modes of vibration. The scaling function also suggests a power-law divergence of $\left<\Psi\right>$ on approaching the unjamming transition, which can be derived from the established knowledge of the vibrational properties of jammed solids. Moreover, we find that $\left<\Psi\right>$ is anisotropic in the directions perpendicular and parallel to the wall. Together with the fact that $\left<\Psi\right>$ is an increasing function of $r$, we are able to predict and understand well the special relaxation dynamics of confined supercooled liquids, including the decrease of the structural relaxation time away from the wall and the anisotropy of the relaxation time in the directions parallel and perpendicular to the wall. Our study thus provides another evidence suggesting that behaviors of supercooled liquids can be viewed from the perspective of amorphous solids at $T=0$ \cite{wanglj,wanglj_jamming,nie_nc}.

In this work, we are only concerned about systems confined effectively by two parallel walls. It remains a question whether the scaling relations found here rely on the shape of the the walls. For instance, the scaling exponents may not be necessarily the same in couette geometry confined by a circular wall. Moreover, here we use the same potential for particle-particle and particle-wall interactions. The softness of the particle-wall interaction is another tunable parameter affecting the behaviors of confined systems. To have a more thorough picture, the shape, the softness, and even the roughness of the walls need to be concerned in follow-up studies.

Although calculable from the normal modes of vibration, $\Psi$ has its definite physical meaning, which is the response of the mean square particle displacement to the increase of temperature in the harmonic approximation. For glassy states in which particles are trapped, $\Psi$ can be approximated by the local Debye-Waller factor. Because the local Debye-Waller factor can be directly measured in for example colloidal experiments, our study proposes an experimentally accessible approach to probe the critical length scale and jamming criticality predicted in simulations.

In previous studies, we have demonstrated the significance of $\Psi$ in characterizing the structural disorder and mechanical susceptibility of amorphous solids \cite{tonghua_pre} and the interplay between jamming criticality and marginal stability in jammed solids \cite{tonghua_prl}. Here we extend the application of $\Psi$ and reveal further its important role in understanding amorphous solids in confinement. We expect to see its more power in the deeper understanding of the key issues of amorphous solids in future studies, including marginal stability, anharmonicity, jamming criticality, and their relations.

\addcontentsline{toc}{chapter}{Acknowledgment}
\section*{Acknowledgment}

This work was supported by the National Natural Science Foundation of China under Grant No.~11734014. We thank the Supercomputing Center of University of Science and Technology of China for the computer time.


\begin{thebibliography}{99}

\item[$^{\dagger}$]huatong@sjtu.edu.cn

\item[$^*$]ningxu@ustc.edu.cn

\bibitem{debenedetti} Debenedetti P G and Stillinger F H {2001 \textit{Nature} \textbf{410} 259}

\bibitem{liu_nagel} Liu A J and Nagel S R {1998, \textit{Nature} \textbf{396} 21}

\bibitem{berthier_rmp} Berthier L and Biroli G {2011, \textit{Rev. Mod. Phys.} \textbf{83} 587}

\bibitem{ohern} O'Hern C S, Silbert L E, Liu A J and Nagel S R {2003, \textit{Phys. Rev. E} \textbf{68} 011306}

\bibitem{liu_review} Liu A J and Nagel S R {2010, \textit{Annu. Rev. Condens. Matter Phys.} \textbf{1} 347}

\bibitem{hecke} van Hecke M {2010, \textit{J. Phys.: Condens. Matter} \textbf{22} 033101}

\bibitem{xu_review1} Xu N {2011, \textit{Front. Phys.} \textbf{6} 109}

\bibitem{xu_review2} Xu N {2019, \textit{Chinese J. Polym. Sci.} \textbf{37} 1065}

\bibitem{ikeda} Ikeda A, Berthier L and Sollich P {2012, \textit{Phys. Rev. Lett.} \textbf{109} 018301}

\bibitem{parisi} Parisi G and Zamponi F {2010, \textit{Rev. Mod. Phys.} \textbf{82} 789}

\bibitem{charbonneau} Charbonneau P, Kurchan J, Parisi G, Urbani P and Zamponi F {2017, \textit{Annu. Rev. Condens. Matter Phys.} \textbf{8} 265}

\bibitem{silbert} Silbert L E, Liu A J and Nagel S R {2005, \textit{Phys. Rev. Lett.} \textbf{95} 098301}

\bibitem{wyart1} Wyart M {2005, \textit{Ann. Phys. Fr.} \textbf{30} 1}

\bibitem{wyart2} Wyart M, Silbert L E, Nagel S R and Witten T A {2005, \textit{Phys. Rev. E } \textbf{72} 051306}

\bibitem{wyart3} Wyart M, Nagel S R and Witten T A. {2005, \textit{Europhys. Lett.} \textbf{72} 486}

\bibitem{xu1} Xu N, Vitelli V, Wyart M, Liu A J and Nagel S R {2009, \textit{Phys. Rev. Lett.} \textbf{102} 038001}

\bibitem{wangxp} Wang X, Zheng W, Wang L and Xu N {2015, \textit{Phys. Rev. Lett.} \textbf{114} 035502}

\bibitem{nie} Nie Y, Tong H, Liu J, Zu M and Xu N {2017, \textit{Front. Phys.} \textbf{12} 126301}

\bibitem{widmer} Widmer-Cooper A, Perry H, Harrowell P and Reichman D R {2008, \textit{Nat. Phys.} \textbf{4} 711}

\bibitem{chenke} Chen K, Manning M L, Yunker P J, Ellenbroek W G, Zhang Z, Liu A J and Yodh A G {2011, \textit{Phys. Rev. Lett.} \textbf{107} 108301}

\bibitem{wanglj} Wang L and Xu N {2014, \textit{Phys. Rev. Lett.} \textbf{112} 055701}

\bibitem{tonglj} Tong H and Tanaka H {2020, \textit{Phys. Rev. Lett.} \textbf{124} 225501}

\bibitem{berthier_attr} Berthier L and Tarjus G {2009, \textit{Phys. Rev. Lett.} \textbf{103} 170601}

\bibitem{tonghua_prl} Tong H, Hu H, Tan P, Xu N and Tanaka H {2019, \textit{Phys. Rev. Lett.} \textbf{122} 215502}

\bibitem{tonghua_pre} Tong H and Xu N {2014, \textit{Phys. Rev. E} \textbf{90} 010401(R)}

\bibitem{chenke_s2} Yang X, Liu R, Yang M, Wang W-H and Chen K {2016, \textit{Phys. Rev. Lett.} \textbf{116} 238003}

\bibitem{chenke_psi} Yang X, Tong H, Wang W-H and Chen K {2019, \textit{Phys. Rev. E} \textbf{99} 062610}

\bibitem{tong_nrp} Tanaka H, Tong H, Shi R and Russo J {2019, \textit{Nature Rev. Phys.} \textbf{1} 333}

\bibitem{watanabe} Watanabe K, Kawasaki T and Tanaka H {2011, \textit{Nat. Mater.} \textbf{10} 512}

\bibitem{scheidler} Scheidler P, Kob W and Binder K {2002, \textit{Europhys. Lett.} \textbf{59} 701}

\bibitem{cao} Cao C, Huang X, Roth C B and Weeks E R {2017, \textit{J. Chem. Phys.} \textbf{147} 224505}

\bibitem{hocky} Hocky G M, Berthier L, Kob W and Reichman D R {2014, \textit{Phys. Rev. E} \textbf{89} 052311}

\bibitem{fire} Bitzek E, Koskinen P, Gahler F, Moseler M and Gumbsch P {2006, \textit{Phys. Rev. Lett.} \textbf{97} 170201}

\bibitem{olsson} Olsson P and Teitel S {2007, \textit{Phys. Rev. Lett.} \textbf{99} 178001}

\bibitem{ellenbroek} Ellenbroek W G, Somfai E, van Hecke M and van Saarloos W {2006, \textit{Phys. Rev. Lett.} \textbf{97} 258001}

\bibitem{wanglj_jamming} Wang L and Xu N {2013, \textit{Soft Matter} \textbf{9} 2475}

\bibitem{graves} Graves A L, Nashed S, Padgett E, Goodrich C P, Liu A J and Sethna J P {2016, \textit{Phys. Rev. Lett.} \textbf{116} 235501}

\bibitem{liuhao} Liu H, Xie X and Xu N {2014, \textit{Phys. Rev. Lett.} \textbf{112} 145502}

\bibitem{goodrich} Goodrich C P, Liu A J and Nagel S R {2012, \textit{Phys. Rev. Lett.} \textbf{109} 095704}

\bibitem{liaoqy} Liao Q and Xu N {2018, \textit{Soft Matter} \textbf{14} 853}

\bibitem{lerner1} Lerner E, During G and Bouchbinder E {2016, \textit{Phys. Rev. Lett.} \textbf{117} 035501}

\bibitem{lerner2} Lerner E and Bouchbinder E {2017, \textit{Phys. Rev. E} \textbf{96} 020104(R)}

\bibitem{mizuno} Mizuno H, Shiba H and Ikeda A {2017, \textit{Proc. Natl. Acad. Sci. USA} \textbf{114} E9767}

\bibitem{charbonneau1} Charbonneau P, Corwin E I, Parisi G, Poncet A and Zamponi F {2016, \textit{Phys. Rev. Lett.} \textbf{117} 045503}

\bibitem{xu2} Xu N, Liu A J and Nagel S R {2017, \textit{Phys. Rev. Lett.} \textbf{119} 215502}

\bibitem{xu3} Xu N, Vitelli V, Liu A J and Nagel S R {2010, \textit{Europhys. Lett.} \textbf{90} 56001}

\bibitem{nie_nc} Nie Y, Liu J, Guo J and Xu N {2020, \textit{Nat. Commun.} \textbf{11} 3198}

\end{thebibliography}
\end{document}